\documentclass{PoS}
\usepackage{multicol,float, subcaption}
\usepackage{graphicx}
\usepackage{enumitem}
\usepackage{gensymb}
\graphicspath{ {./images/} }
\title{A Method for an Untriggered, Time-Dependent, Source-Stacking Search for Neutrino Flares }

\ShortTitle{Multiflare Stacking}

\author{
The IceCube Collaboration$^{\dagger}$\\
The IceCube Collaboration\footnote{For collaboration list, see PoS(ICRC2019) 1177.}\\
{\itshape \href{http://icecube.wisc.edu/collaboration/authors/icrc19_icecube}{http://icecube.wisc.edu/collaboration/authors/icrc19\_icecube}}\
E-mail: \email{wluszczak@icecube.wisc.edu, jbraun@icecube.wisc.edu}
}

\abstract{

Recent results from IceCube regarding TXS 0506+056 suggest that it may be useful to test the hypothesis of multiple neutrino flares, where each flare is not necessarily accompanied by a corresponding gamma-ray flare. An untriggered, time-dependent, source-stacking search would be optimal for testing this hypothesis, however such an analysis has yet to be applied to the full duration of IceCube data. Here, we discuss one possible way of constructing such an analysis, with a global test statistic obtained by summing the individual test statistics associated with each signal-like flare in the sample. This has the additional advantage of assessing the time structure of each source when running the analysis over a source catalog. We show that, for a signal consisting of many small flares, this style of analysis represents a significant increase in discovery potential over both the existing single-flare fit and the time-integrated stacking methods. Potential source catalogs are examined in combination with this method, including the possibility of a ``self-triggered'' catalog consisting of the locations of the highest energy northern sky events in the IceCube sample. 
\\

\vspace{4mm}
{\bfseries Corresponding authors:}
\speaker{William Luszczak}$^{1}$, Jim Braun$^{1}$, Albrecht Karle$^{1}$\\
{$^{1}$ \itshape Dept. of Physics and Wisconsin IceCube Particle Astrophysics Center, University of Wisconsin, Madison, WI 53706, USA}\\

}

\FullConference{36th International Cosmic Ray Conference -ICRC2019-\\
		July 24th - August 1st, 2019\\
		Madison, WI, U.S.A.}

\begin{document}
\section{Introduction}\label{sec:info}

Recent results from the IceCube Neutrino Observatory suggest the possibility of multiple untriggered neutrino flares from the same object \cite{Aartsen:singleflare}\cite{Aartsen:multimessenger}. The method used in \cite{Aartsen:singleflare} tests the hypothesis of multiple flares for TXS 0506+056 by combining the single, most significant flares found over different pre-defined data taking periods. While this method works well in the case that the data is dominated by a single, significant neutrino flare, it is not sensitive to the case of multiple, similarly-sized flares produced by the same source. Here, we present a method specifically tailored to testing the hypothesis that a particular source flares many (>1) times. We will refer to this new method as "multiflare stacking". The primary difference between multiflare stacking and the existing methods \cite{Braun:sfmethod}, is that multiflare stacking attempts to fit \textit{all} flares in the sample, not merely the largest flare. This method is particularly well suited for testing the hypothesis of many, moderate-to-weakly flaring sources.

\section{Description of the method}

Conceptually, multiflare stacking can be broken down into the following steps:

\begin{enumerate}
    \item Create a set of "test windows", defined by a start and stop time. This is done by taking all possible pairs of events that pass a pre-defined weight threshold. 
    \item Calculating a test statistic associated with each test window, describing the hypothesis that that particular window was a flare produced by the source. Discard all windows with test statistic below some threshold: These windows are "background-like", while the remaining windows (with test statistic > this threshold) are "signal-like".
    \item For the remaining signal-like windows, remove windows that overlap with a window with higher test statistic. Sum the remaining $TS_j$ to obtain a multiflare test statistic, $\widetilde{TS}$, associated with a particular source.
    \item (For the case of considering a source catalog) Do a binomial test to optimize for the number of sources to stack together.
\end{enumerate}

\subsection{Create a set of "test windows", defined by a start and stop time.}
For the purpose of multiflare stacking, we will restrict ourselves to only considering flares which begin and end on an event. This means that for a data sample containing $N$ events, there are at most $N(N-1)/2$ possible flares that can be formed. We will refer to these possible flares (defined solely by a start and stop time) as "test windows". 

\begin{figure}[h]
    \centering
    \includegraphics[width=0.5\textwidth]{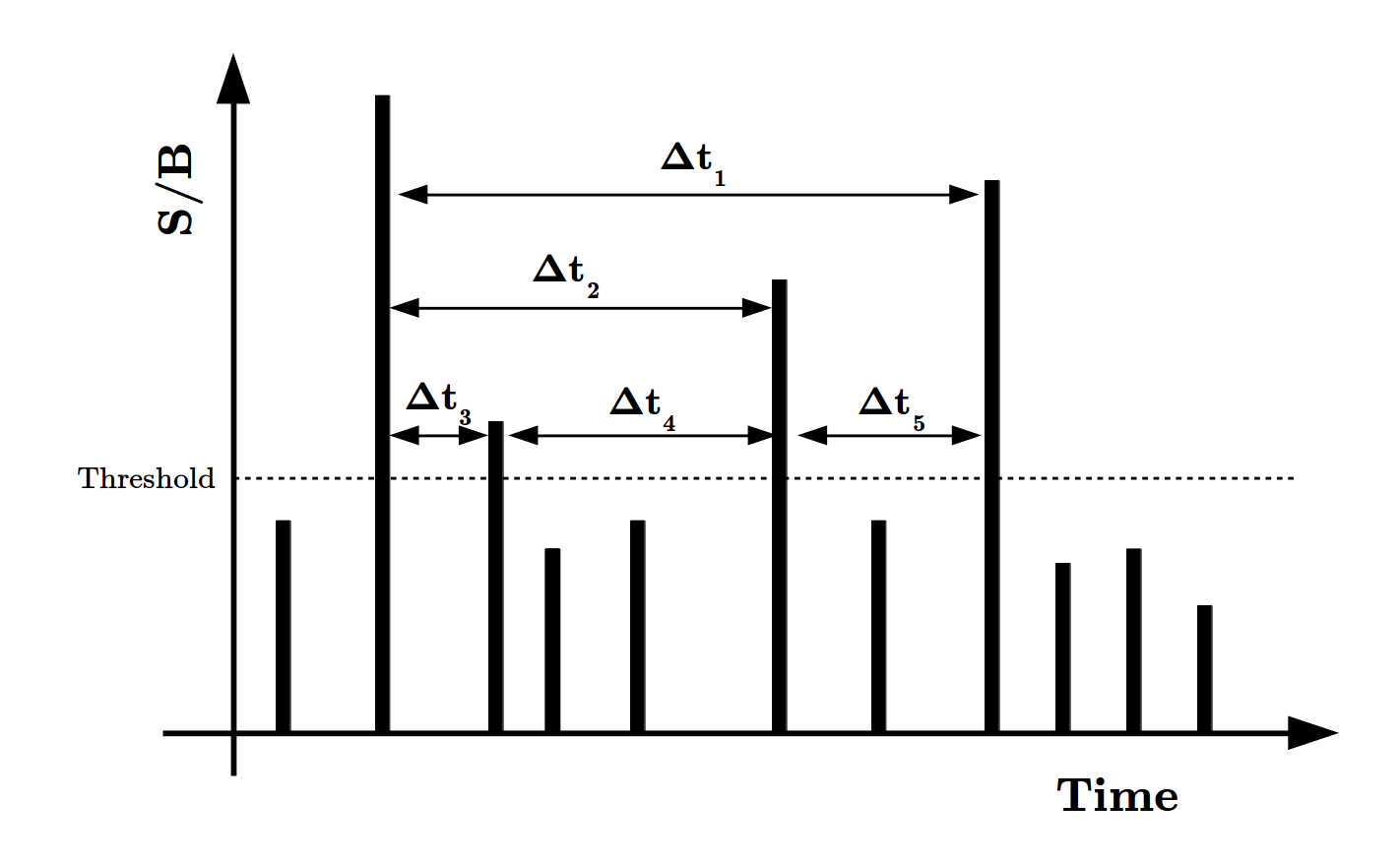}
    \caption{The process of selecting test windows for which a test statistic will be calculated. Test windows may overlap, and windows are only formed from events passing the $S/B$ threshold. Note that while the $S/B$ threshold is applied to determine which windows will be considered, \textit{all} events (regardless of $S/B$ value) will be used when calculating a test statistic. }
    \label{fig:fig1}
\end{figure}

In practice, due to the large number of events, steps 2-4 from the above are computationally infeasible if all $N(N-1)/2$ test windows are considered, so we reduce the number of possible windows by applying a cut to the number of events prior to forming test windows: only events with $S/B$ ratio greater than some threshold will be considered for forming test windows. Here, $S/B$ refers to the ratio of the spatial and energy components of the signal and background PDFs defined in equations 4 and 8 of \cite{Braun:sfmethod}. This significantly reduces the number of test windows to a more manageable number. This process is shown in Figure ~\ref{fig:fig1}. This $S/B$ event selection is only used for the test window determination. For the actual analysis all $N$ events will be used. 

\subsection{Calculating a test statistic associated with each test window}
For each of the test windows defined in section 2.1, we calculate a flare test statistic ($TS_{j}$ associated with that specific flare hypothesis (that the flare starts at the time of the first seed event, and ends at the time of the second seed event). We identify individual test windows with the index $j$. This flare test statistic is of the form presented in section 3 of \cite{Braun:sfmethod}, with the modification that we use a box-shaped flare hypothesis instead of a gaussian. . Signal PDFs ($S_{ij}$ where the index ${i}$ indicates the event) are defined as in equations 2.1-2.3. Events (identified by the index $i$) are defined by a arrival direction ($\vec{r}_i$), energy ($E_i$), and time ($t_i$). The signal PDF is composed of a spatial term ($R_{ij}(\vec{r}_i)$), defined by the location $\vec{r}_i$ and event angular error $\sigma_{i}$. The energy PDF, $\mathcal{E}(E_{ij}|\gamma_{j})$, describes the probability of obtaining an event energy ($E_i$) from a signal event produced by a flare with spectrum $\gamma_j$, and is a function both of $\gamma_j$ and detector effective area. Note that we include a $j$ index with the flare spectrum, $\gamma_j$, as this method will fit the spectrum of each flare individually. In the definition of the temporal PDF (equation 2.3), $H(t)$, is the Heaviside function, $\Delta T_j$ is the test window duration, and $t_j^{start}$ and $t_j^{end}$ are the start and end time of the $j$th test window. 

\begin{equation}
S_{ij} = R_{ij}(\vec{r_i}) \times \mathcal{E}(E_i|\gamma_j) \times \mathcal{T}_{ij}(t_{i})
\end{equation}
\begin{equation}
R_{ij}(\vec{r_i}) = \frac{1}{2 \pi \sigma_{i}^2}exp[\frac{-r_i^2}{2\sigma_i^2}]
\end{equation}
\begin{equation}
\mathcal{T}_{ij}(t_i) = \frac{H(t_{j}^{end}-t_i)H(t_i-t_j^{start})}{\Delta T_j}
\end{equation}

The background PDF is defined in equation 2.4. Similar to the signal PDFs, the background energy and spatial PDFs are functions of the detector effective area, while the temporal background PDF, $\frac{1}{\Delta T_{data}}$, is simply a uniform distribution over the entire data taking period.

\begin{equation}
B_{i} = R_{bg}(\vec{r_i})\mathcal{E}(E_i|Atm_{\nu}) \frac{1}{\Delta T_{data}}
\end{equation}

We can then define our test window likelihood and associated test statistic, shown in equations 2.5 and 2.6. Here, $N$ refers to the total number of events in the sample (without the $S/B$ requirement mentioned above), and $n_{sj}$ is the number of signal events contained within the $j$th flare. The flare likelihood (equation 2.5) can be minimized with respect to $n_{sj}$ and $\gamma_j$ to obtain the flare test statistic in equation 2.6, where $\hat{n}_{sj}$ and $\hat{\gamma}_j$ refer to the best-fit values resulting from minimizing equation 2.5. This test statistic describes the hypothesis that the $j$th flare is in fact a signal flare. Unlike the standard method, we do not minimize our likelihood with respect to the flare parameters $\Delta T_j$, $t_j^{end}$ or $t_j^{start}$, as instead we seek to describe the hypothesis of a flare with these parameters fixed to specific values. 

\begin{equation}
    \mathcal{L}_j(n_{sj}, \gamma_j) = \prod_{i=1}^{N}(\frac{n_{sj}}{N}S_{ij} + (1-\frac{n_{sj}}{N})B_{i})
\end{equation}

\begin{equation}
    TS_j = -2 \log \left [\frac{\Delta T_{data}}{\Delta T_j} \times \frac{\mathcal{L}_j(n_{sj}=0))}{\mathcal{L}_j(n_{sj}=\hat{n}_{sj}, \gamma_j=\hat{\gamma}_j)}\right]
\end{equation}

The $\Delta T_{data}/ \Delta T_j$ term is intended to correct for the fact that there are many small windows that formed in section 2.1, so we penalize their test statistic to account for the additional trials associated with short test windows. Note that while test windows are defined by events that pass the threshold in step 1, the $i$ index in equation 2.5 runs over \textit{all} events, not only those which pass the $S/B$ cut. 
 
This process can then be repeated for every test window defined in section 2.1. The end result of this is a list of flares (start/stop times), and corresponding $TS_j$'s. We then discard all windows with $TS_j < 0$. These windows are "background-like", while the remaining windows, with $TS_j > 0$ are "signal-like". For the remaining steps we will work only with the list of "signal-like" windows.

\subsection{Removing overlapping windows}
\begin{figure}[t!]
    \centering
    \caption{Multiflare stacking applied to a single source}
    \begin{subfigure}[t]{0.49\textwidth}
        \centering
        \includegraphics[width=0.99\textwidth]{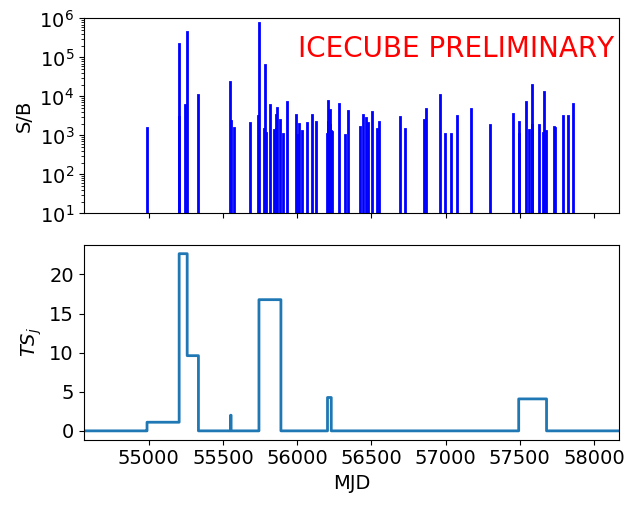}
        \subcaption{Top: Events that pass the $S/B$ threshold cut for generating windows, mentioned in section 2.1, for a source located at (ra, dec) = ($77.45^{\circ}$, $5.61^{\circ}$). This source contains three injected flares, centered at MJD=55246.3, 55807.4, and 57632.7. Bottom: The resultant flare curve calculated by applying multiflare stacking to this source. The single flare method fits only the flare at MJD = 55246.3 (with TS = 22.66), while the multiflare method fits all flares together, with a global test statistic of $\widetilde{TS}$ = 60.52.}
    \end{subfigure}
    ~
    \begin{subfigure}[t]{0.49\textwidth}
        \centering
        \includegraphics[width=0.99\textwidth]{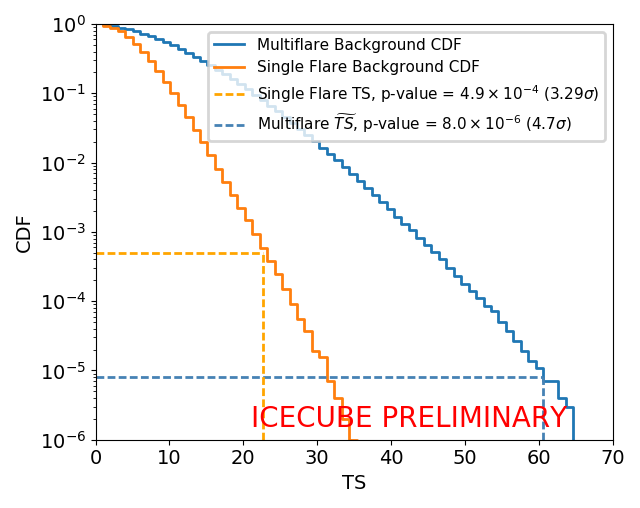}
        \subcaption{The background test statistic distributions for a single source, located at declination = $5.61^{\circ}$. The background multiflare test statistic distribution is shown in blue, while the single flare method is shown in orange. The vertical lines represent the test statistics associated with the flare curve in figure 2a. The single flare p-value for this source is $4.9 \times 10^{-4}$ (3.29$\sigma$), while the multiflare p-value is $8.0 \times 10^{-6}$ ($4.7\sigma$).}
    \end{subfigure}
    \label{fig:fig2}
\end{figure}

Individual events should not be able to contribute to more than one flare. To account for this, we employ the following decorrelation procedure: test windows are ordered by descending $TS_j$ (calculated above). Iterating down the list, discard all test windows that overlap in time with a different test window with a larger $TS_j$. The result of this step is a list of completely decorrelated (non-overlapping) test windows. This list can be interpreted as a "neutrino flare curve", describing the set of neutrino flares associated with a particular source location. Figure ~\ref{fig:fig2} shows an example flare curve generated by this procedure.

We can now define the "global" test statistic, $\widetilde{TS}$ to be the sum of all the $TS_j$ remaining after this decorrelation procedure, as in equation 2.7, where the sum in equation 2.7 is only over non-overlapping flares. The best fit value for the number of signal events, $\tilde{n}_s$, can be obtained in a similar manner by simply summing up the best fit $\hat{n}_s$ associated with all the non-overlapping flares.  
\begin{equation}
     \widetilde{TS} = \sum_{j, TS_j>0}TS_j
     \quad
     \widetilde{n}_s = \sum_{j, TS_j>0}\hat{n}_{sj}
\end{equation}

\begin{figure}[h!]
    \centering
    \begin{subfigure}[h]{0.49\textwidth}
        \centering
        \includegraphics[width=0.99\textwidth]{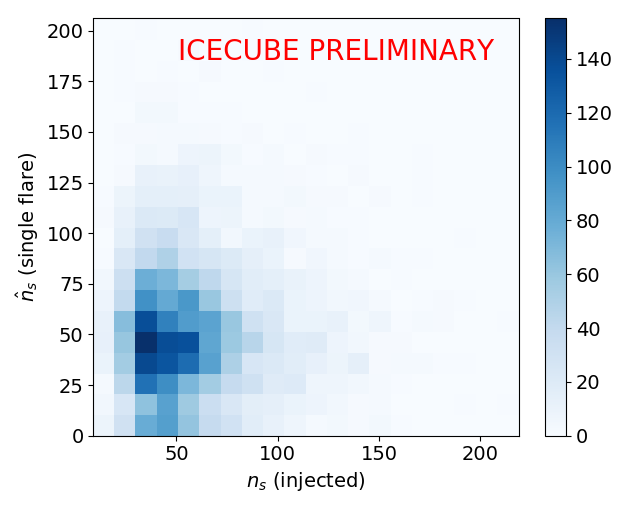}
    \end{subfigure}
    ~
    \begin{subfigure}[h]{0.49\textwidth}
        \centering
        \includegraphics[width=0.99\textwidth]{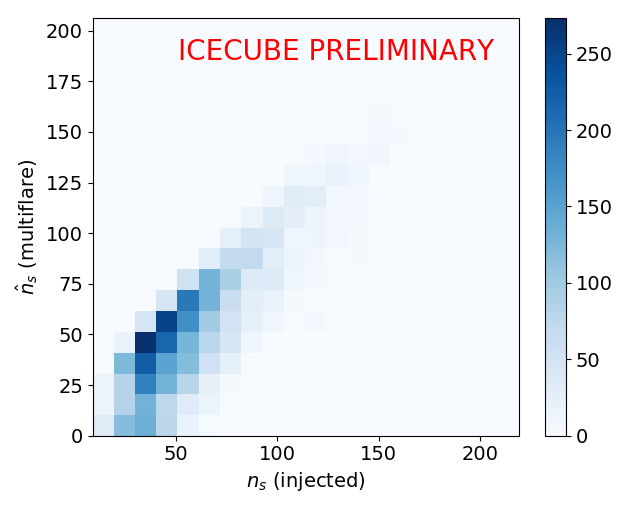}
    \end{subfigure}
    \caption{Fit versus injected values for the number of signal events ($n_s$), for both the existing single flare method (left) and the multiflare method presented in these proceedings (right). In both cases a signal composed of many flares was injected, with flare intensities drawn according to the power law distribution introduced in section 3. Unsurprisingly, multiflare stacking is much more capable of recovering the number of injected signal events for a signal of this form. }
    \label{fig:fig3}
\end{figure}

Figure ~\ref{fig:fig3} shows the best fit $\tilde{n}_s$ versus the injected value for both this method and the traditional single flare fit. For a signal composed of many flares, multiflare stacking recovers the injected signal much more accurately.

\subsection{Procedure to treat multiple sources from a catalog}
We can easily extend this process to a source catalog, rather than a single source. The global test statistic is simply given by equation 2.8, where here the sum runs over all sources in the source catalog. 
\begin{equation}
     \widetilde{TS}_{all} = \sum_{sources}\widetilde{TS}
\end{equation}

For large source catalogs, if only a few sources are "truly" flaring, they may be overwhelmed by the large number of background sources. This can be remedied by "binomial stacking" of sources. Instead of stacking the multiflare test statistic for all sources (equation 2.8), we instead employ the following procedure to optimize the number of sources to stack ($k$). For a given data set, the sources are ordered by their multiflare test statistic, $\widetilde{TS}$. A p-value for $k=1,2,3,...N_{srcs}$ is calculated, and subsequently the $k$ that produces the minimum p-value is selected. An additional trial factor is attached from a newly created background test statistic distribution, corresponding to the number of stacked sources, similar to the process described in \cite{Erin:2019icrc_blazars}.  The additional power obtained by stacking in this manner outweighs the penalty associated with the trial factor, resulting in an increased discovery potential (see section 4). 

\begin{equation}
     \widetilde{TS}_{all} = \sum_{m=0}^{k_{best}}\widetilde{TS}_m
\end{equation}

\section{Characterizing an Injected Signal}

In order to test the performance of multiflare stacking, we require a description of the intensity distribution of observed flares (i.e. how many events are in each flare, and how many flares there are total). For flares of intensity $I$, the number of flares with that intensity $N(I)$ can be described by a power law with a lower cutoff (equation 3.1). This is primarily motivated by the distribution of gamma ray flares found in \cite{Fermi:fava}. Note, however, that this is a \textit{descriptive} power law, not a \textit{predictive} power law. The parameters here are intended as a flexible way to describe a wide array of what the distribution of flares may actually look like. We make no claims as to what these parameters actually \textit{are} beyond the existing limits calculated in section 4. 

\begin{equation}
    N(I) = A_m(\frac{\alpha-1}{I_o})(\frac{I}{I_o})^{-\alpha}
\end{equation}

This power law is specified by three parameters:

\begin{itemize}[nosep]
    \item $A_m$ describes the overall power law normalization (how many flares exist in the sample)
    \item $I_o$ describes the lower intensity cutoff. This is a parameter that scales with the flux, but roughly corresponds to the minimum flare size.
    \item $\alpha$ describes the number of larger versus smaller flares. A larger value of $\alpha$ corresponds to proportionally more smaller flares, while a smaller $\alpha$ will produce more larger flares.
\end{itemize}

\section{Catalogs and Discovery Potential}
In this section we will explore the 3$\sigma$ discovery potential of multiflare stacking applied to two different catalogs. We define the 3$\sigma$ discovery potential to be the flare intensity distribution normalization ($A_m$) required for 50\% of the injected signal multiflare test statistics to be higher than the 3$\sigma$ threshold calculated from the background multiflare test statistic distribution. Under this construction, the 3$\sigma$ discovery potential is a surface in the space spanned by possible values of $A_m$, $\alpha$ and $I_o$. For plots in this section, we have chosen to show slices of this space with fixed $\alpha$ ($\alpha=3$), plotting the discovery potential $A_m$ as a function of $I_o$. Since we have yet to observe a flaring neutrino source, the allowed values of $\alpha$ are unconstrained, we have simply chosen a specific value to illustrate a region where multiflare stacking is relevant. This is purely for the purposes of demonstration, as this method has significant discovery potential for a wide range of injected values of $\alpha$.

\subsection{Catalog 1: The Top 500 Flux Fermi 3LAC Blazars}

Due to previous results of time-integrated stacking searches applied to Fermi 2LAC blazars \cite{Aartsen:Blazar_tint}, it may be beneficial to test the hypothesis of multiple flares for the catalog of Fermi 3LAC blazars. We select the top 500 flux sources in this catalog purely due to computational limitations, but otherwise include no additional cuts. 

\begin{figure}[h!]
    \centering
    \begin{subfigure}[t]{0.49\textwidth}
        \includegraphics[width=0.99\textwidth]{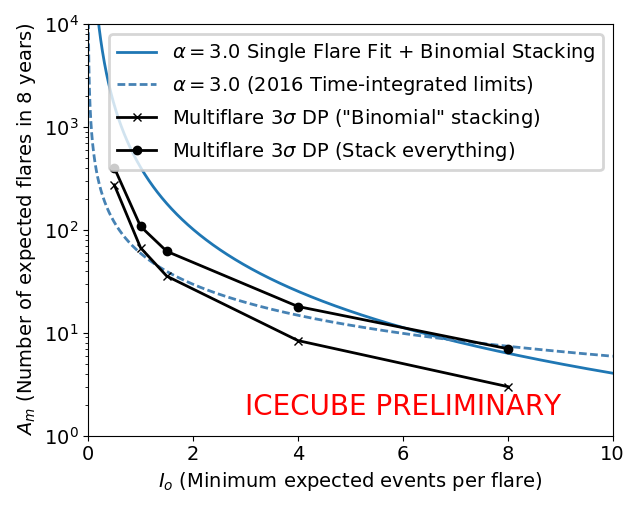}
        \caption{The 3$\sigma$ multiflare discovery potential for a catalog of Fermi 3LAC blazars. The dashed line corresponds to the limits set in \cite{Aartsen:Blazar_tint}.}
        \label{fig:fig5}
    \end{subfigure}
    ~
    \begin{subfigure}[t]{0.49\textwidth}
        \includegraphics[width=0.99\textwidth]{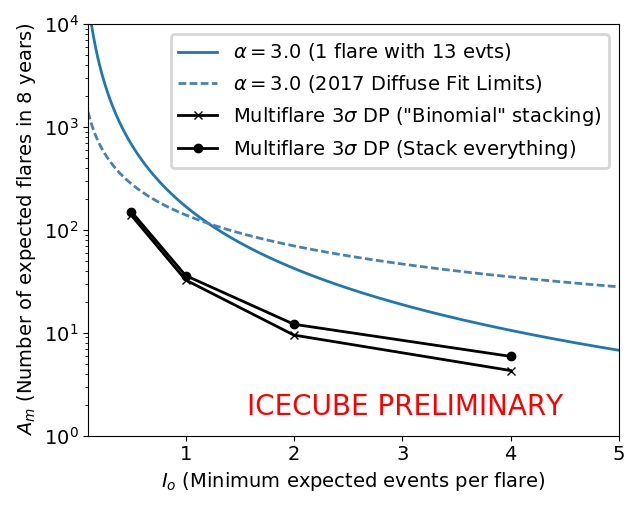}
        \caption{The existing limits and 3$\sigma$ discovery potential for the "self-triggered" catalog described in section 4.2. The dashed line in this plot refers to the diffuse fit done in \cite{Aartsen:Diffuse_fit}.}
        \label{fig:fig6}
    \end{subfigure}
    \caption{Multiflare discovery potential. Injected flare durations are fixed to 100 days, and the injected events follow an $E^{-2.0}$ spectrum. Note that the discovery potential improves a small, but significant amount with the inclusion of binomial-style stacking (section 2.4). In both figures, the solid blue line corresponds to the observation of a single, 100 day flare with 13 events.}
\end{figure}

The currently existing limits for this catalog are plotted in blue in figure  ~\ref{fig:fig5}, and can be calculated analytically by integrating equation 3.1. The "single flare limits" correspond to performing a binomial test on the Fermi 3LAC blazar catalog (a 3 sigma discovery corresponds to roughly 5 pre-trial 3-sigma injections), as in ~\cite{Erin:2019icrc_blazars}. The "time integrated limits" correspond to the condition that there are not more events in the sample than are allowed by the existing time-integrated flux limits reported in \cite{Aartsen:Blazar_tint}.

Figure ~\ref{fig:fig5} shows the 3$\sigma$ discovery potential for applying multiflare stacking to this source catalog. Note that the binomial-style stacking method proposed in section 2.4 significantly improves the discovery potential, improving it beyond the currently existing limits set by \cite{Aartsen:Blazar_tint}.  Also note that for large values of $I_o$, the discovery potential of the single flare fits is expected to exceed that of multiflare stacking, since this region corresponds to having only 1-2 large flares in the sample. For very small values of $I_o$, the time integrated method becomes superior, since in this region "flares" are often merely a single event. Multiflare stacking is primarily sensitive to the middle region, where there are a moderate amount of medium-sized flares.

\subsection{Catalog 2: High-Energy IceCube Events}

Based on the diffuse fits to the atmospheric and astrophysical spectrum in \cite{Aartsen:Diffuse_fit}, we can develop a "self-triggered" catalog of source locations consisting of the highest energy events in the sample. This is similar in spirit to the catalog considered in \cite{Karl:2019icrc_selftriggered}. For this catalog, we test the locations of all events in the sample with energy proxy > 200 TeV. This threshold is chosen because above 200 TeV, we expect 50\% of events to be astrophysical in origin \cite{Aartsen:Diffuse_fit}. These "source" events are removed from the sample prior to calculating a test statistic to avoid signal contamination.

This catalog consists of 32 high-energy events, though notably does not include IC170922A, as this data sample only extends up to 2016. Events are well localized, with the majority of events having reconstructed angular error < $1^{\circ}$, so these locations can be considered to be point-like source candidates.

The discovery potential for this catalog is shown in figure ~\ref{fig:fig6}. The benefit of the binomial test is smaller in this case, due to the size of the source catalog, however multiflare stacking still represents a significant improvement over both the single flare and time-integrated methods.

\section{Concluding Remarks}

We have presented a method for testing the hypothesis that neutrino sources flare multiple times ("multiflare stacking"). This is done by combining information from fits to all flares in the sample. Using an injected signal characterized by a power law, we explored the discovery potential for this method for two source catalogs: Fermi 3LAC blazars and a "self-triggered" catalog composed of the locations of high energy IceCube events observed between 2009 and 2016. For both of these catalogs, multiflare stacking represents a significant improvement over existing methods for a signal composed of many moderately-sized flares.

\bibliographystyle{ICRC}
\bibliography{references}
\end{document}